\documentclass[aip,graphicx,amsmath,amssymb,reprint]{revtex4-1}
\usepackage{graphicx}
\usepackage{dcolumn}
\usepackage{bm}
\usepackage[utf8]{inputenc}
\usepackage[T1]{fontenc}
\usepackage{mathptmx}
\usepackage{amsmath}
\usepackage{amsfonts}
\usepackage{amssymb}
\usepackage{isomath}
\usepackage[usenames]{color}
\usepackage{colortbl}
\draft 

\begin{document}


\title{Electrically tunable detector of THz-frequency signals based on an antiferromagnet} 



\author{A.~Safin}
\email{arsafin@gmail.com}
\affiliation{Kotel’nikov Institute of Radioengineering and Electronics, Russian Academy of Sciences, 125009 Moscow, Russia}%
\affiliation{National Research University “Moscow Power Engineering Institute,” 111250 Moscow, Russia}%

\author{V.~Puliafito}
\email{vpuliafito@unime.it}
\affiliation{Department of Engineering, University of Messina, I-98166 Messina, Italy}%

\author{M.~Carpentieri}%
\affiliation{Department of Electrical and Information Engineering, Politecnico di Bari, I-70125 Bari, Italy}%

\author{G.~Finocchio}
\affiliation{Department of Mathematical and Computer Sciences, Physical Sciences and Earth Sciences, University of Messina, I-98166 Messina, Italy}%

\author{S.~Nikitov}
\affiliation{Kotel’nikov Institute of Radioengineering and Electronics, Russian Academy of Sciences, 125009 Moscow, Russia}%
\affiliation{Moscow Institute of Physics and Technology, Dolgoprudny, Moscow Region 141700, Russia}%
\affiliation{Laboratory ”Metamaterials”, Saratov State University, 83 Astrakhanskaya street, Saratov, 410012, Russia}%

\author{P.~Stremoukhov}
\affiliation{Kotel’nikov Institute of Radioengineering and Electronics, Russian Academy of Sciences, 125009 Moscow, Russia}%
\affiliation{Moscow Institute of Physics and Technology, Dolgoprudny, Moscow Region 141700, Russia}%
\affiliation{FELIX Laboratory, Radboud University, 6525 AJ Nijmegen, The Netherlands}%

\author{A.~Kirilyuk}
\affiliation{Kotel’nikov Institute of Radioengineering and Electronics, Russian Academy of Sciences, 125009 Moscow, Russia}%
\affiliation{FELIX Laboratory, Radboud University, 6525 AJ Nijmegen, The Netherlands}%

\author{V.~Tyberkevych}
\affiliation{Oakland University, Rochester, Michigan 48309, USA}%

\author{A.~Slavin}
\affiliation{Oakland University, Rochester, Michigan 48309, USA}%


\date{\today}

\begin{abstract}

 A concept of an electrically tunable resonance detector of THz-frequency signals based on antiferromagnetic/heavy metal (AFM/HM) hetero-structure is proposed. The conversion of a THz-frequency input signal into DC voltage is done using the inverse spin Hall effect in an (AFM/HM) bilayer. An additional bias DC current in the HM layer can be used to vary the effective anisotropy of the AFM, and, therefore, to tune the AFMR frequency. The proposed AFM/HM hetero-structure works as a resonance-type quadratic detector which can be tuned by the bias current in the range of at least 10 percent of the AFMR frequency, and our estimations show that the sensitivity of this detector could be comparable to that of modern detectors based on the Schottky, Gunn or graphene-based diodes.
\end{abstract}

 \maketitle 

There is a growing interest in the development of tunable oscillators and detectors operating in a terahertz (THz) frequency range. Antiferromagnetic (AFM) materials have natural resonance frequencies of spin excitations (antiferromagnetic resonance or AFMR) lying in this frequency range. Some AFM materials can operate at room temperatures, they do not require any bias magnetic field, and can be tuned by changing their anisotropy. These properties make AFM materials very attractive for use in THz-frequency signal processing devices.The resonance detectors of terahertz (THz) frequency signals have a great potential for use in non-destructive testing, security and telecommunication technologies~\cite{pawar2013terahertz, federici2005thz, sizov2010thz, ferguson2002materials} since the THz-frequency radiation has a relatively large penetration depth, being, at the same time, non-ionizing. However, generation and resonance detection of signals with frequencies lying in the so-called "THz-gap" (from 0.1 to 10 THz) is rather difficult due to the rarity of naturally existing resonators operating in this frequency range. Vacuum electronic devices, semiconductor and graphene-based oscillators with frequency  multipliers can generate high-amplitude signals at frequencies of up to several hundred GHz~\cite{lewis2014review}, whereas light-based sources, such as quantum cascade lasers, provide signals with frequencies higher than several THz at room temperature~\cite{williams2008terahertz}, so the THz-gap still exists. Some relief is provided by the oscillators and detectors based on Josephson junctions~\cite{ozyuzer2007emission, hu2010phase, barone1982physics, likharev1986dynamics}, but these devices require cryogenic temperatures for their operation, which creates a significant difficuly in their practical use. Thus, the development of a small and simple room-temperature devices capable of generating and/or receiving resonantly signals in a "THz-gap" is still a significant challenge.

It was suggested previously to use AFM materials  as active layers of THz-frequency oscillators, due to the fact that the strong internal exchange magnetic field existing inside the AFM crystals pushes the frequencies of signals that can be generated in these crystals into the THz-frequency range ~\cite{Cheng2016}.  Also, there are several theoretical papers ~\cite{khymyn2017antiferromagnetic, sulymenko2017terahertz, sulymenko2018terahertz} that suggested the possibility of development of non-resonant continuously tunable current-driven THz-frequency auto-oscillators based on the effect of rotation of the AFM magnetic sublattices tilted by an external DC spin current in the large internal exchange magnetic field existing inside an AFM. It was predicted, that the generation frequency of such AFM/HM - based  auto-oscillators, controlled by the DC bias electric current flowing in the HM  layer, would vary between 0.1–2.0~THz when the bias DC current would be varied between  $10^8$~A/cm$^2$  and $10^9$~A/cm$^2$.

Recently~\cite{gomonay2018narrow, Khymyn2018}, it has been theoretically proposed to use active AFM generators for the detection of external THz-frequency signals via the mechanism of  injection-locking of such a signal to the oscillations generated by a DC-current-driven AFM/HM THz generator.

An alternative way to develop \emph{quasi-passive} AFM/HM-based detectors ~\cite{khymyn2017antiferromagnetic1}is to use the fact that resonance eigenfrequencies of the AFM dynamic modes (standing AFMR modes) lie in the THz frequency range. It has been shown theoretically in ~\cite{khymyn2017antiferromagnetic1}, that a dielectric AFM having bi-axial anisotropy, such as NiO, can be used for the resonance quadratic rectification of a linearly-polarized AC spin current of THz-frequency, and could have a sensitivity in the range of $10^2-10^3$~V/W.

The theoretical estimations of the AFM/HM detector parameters presented in~\cite{khymyn2017antiferromagnetic1} are rather encouraging, but for the practical use of such a detector it is highly desirable to be able to \emph{continuously tune} the resonance (AFMR) frequency of such a device by electric means, which was the main motivation of this work.

It should be noted, that in the GHz frequency range similar quadratic detectors based on the spin-torque magnetic diode (STMD) effect in ferromagnetic tunnel junctions has been investigated both theoretically and experimentally~\cite{Tulapurkar2005, Prokopenko2011}. The operating frequency of the ferromagnetic  STMD is limited by the maximum possible applied bias magnetic field,  and it is practically impossible to increase this frequency above several tens of GHz.

For the detector devices based on an AFM/HM hetero-structure the resonance (AFMR) frequencies are proportional to the square root of the product of the internal exchange and anisotropy magnetic fields (see Eq.(4) below). While the internal exchange magnetic field  is very large (it reaches hundreds of Tesla) and fixed by the strong homogeneous exchange interaction, the AFM anisotropy field can be relatively easily controlled by various external means. In this work, we demonstrate a possibility to control the AFM anisotropy field, and, therefore, the AFMR frequency, by changing a DC bias current in the HM layer of the AFM/HM hetero-structure. To confirm our analytical results on the current-induced AFMR frequency tuning, we performed micromagnetic simulations by solving numerically the Landau-Lifthitz-Gilbert equation with a current-induced term (for more details see our previous work~\cite{puliafito2019micromagnetic}).

\emph{}

 \begin{figure}
	\includegraphics[width=0.45\textwidth]{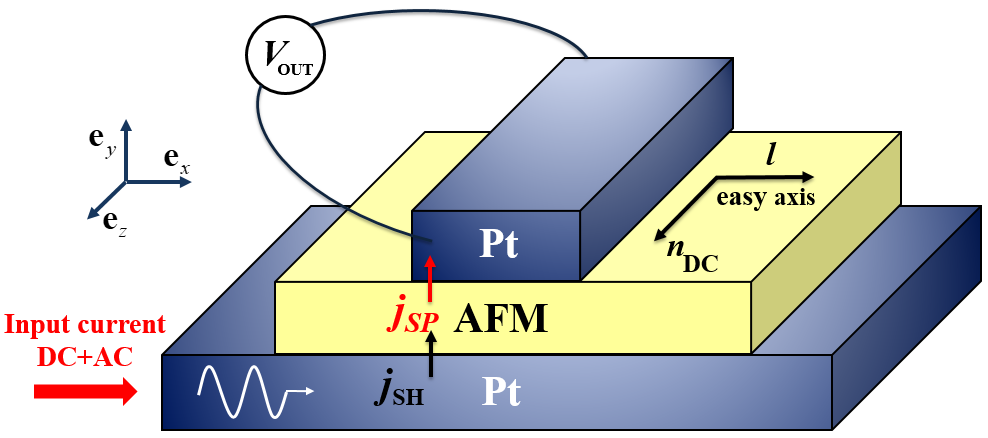}%
	\caption{\label{Fig1} Schematic view of the THz-frequency resonance detector based on the AFM-Pt structure, where ${\bm l}$ is the N\'eel vector oriented along the easy axis ${\bm n}_\text{e}=\bm{e}_\text{x}$ and $V_\text{OUT}$ is the output DC electric voltage. Due to the spin Hall effect input electric current in Pt creates a spin current ${\bm j}_{SH}$, which has both DC and AC components. The polarization of the AC spin current is directed perpendicular to the interface, while the polarization of the DC current ${\bm p}_{DC}={\bm e}_\text{z}$ is oriented in the interface plane and perpendicular to the easy axis. Oscillations of the N\'eel vector cause a spin current ${\bm j}_\text{SP}$ due to the spin-pumping mechanism. Both AC and DC spin-pumping signals are transformed into electric field signals via the inverse spin-Hall effect in the second Pt layer placed on top of the AFM~\cite{Yang2016, Wu2016}.}.
\end{figure}

In this work we consider a THz detector, schematically shown in Fig.1, which consists of an \emph{uniaxial} AFM driven by both DC (bias) and AC (signal) spin currents flowing from the bottom Pt-layer. A spin current of the  density ${\bm j}_{SH}$, produced due to the spin-orbit interaction in the bottom Pt layer, flows into the AFM, and creates DC and AC spin torques with polarizations $\bm{n}_{\text{DC}}$ and $\bm{n}_{\text{AC}}$, respectively. These torques are acting on the magnetic sublattices. The precession of the magnetic sublattices due to the spin-pumping mechanism creates a spin current ${\bm j}_\text{SP}$. This spin current via the inverse spin-Hall effect induces in the top Pt layer (see Fig.1) an electric field in the direction $\bm{e}_\text{z}$ perpendicular to the DC charge current flowing in the $\bm{e}_\text{x}$-direction.
We are interested in the DC part of the electric voltage $V_\text{OUT}$ induced in the top Pt layer between the output contacts which are separated by the distance $L=10\mu$m. Note, that for the  practical implementation of the proposed detector one needs to measure a relatively small rectified DC voltage, proportional to the amplitude of the input AC signal, in the case when the bias DC current could be rather large. Therefore, it is very important, to guarantee that the source supplying the bias DC current is highly stable.
 
 Here, we consider a case when the polarization vector of ${\bm n}_\text{DC}={\bm e}_\text{z}$ of the DC current is oriented perpendicular to the easy axis ${\bm e}_\text{e}$ of the AFM anisotropy, and  the input AC spin current has a circular polarization described by the vector  ${\bm n}_\text{AC}~=~({\bm e}_\text{y}~ \pm~ i {\bm e}_\text{z})/\sqrt{2}$, where two signs correspond to the clockwise and anti-clockwise rotation of the AC spin current polarization in the plane perpendicular to the easy axis ${\bm e}_\text{e}$ of the AFM anisotropy. In should be mentioned, that, as it was was shown in~\cite{khymyn2017antiferromagnetic1}, a \emph{linearly } polarized AC spin current induces a \emph{zero }output DC voltage in an uniaxial AFM, so the circular polarization of the input AC current is critical. An input AC current having circular polarization could, for example, originate from a THz-frequency signal source placed in an EM resonator with a circular polarized magnetic field ~\cite{Jain1997}, or could be obtained using an additional magnetic layer, which creates spin current with circular polarization in the z-y plane, or could be supplied by any other AFM-based THz-frequency oscillator~\cite{khymyn2017antiferromagnetic,khymyn_2019_FiM}.

We describe the AFM magnetization dynamics using the N\'eel vector ${\bm l}~=~({\bm M}_1 - {\bm M}_2)/2M_s$, where ${\bm M}_{1,2}$ are the magnetization vectors of the AFM sublattices, and $M_\text{s}$ is the saturation magnetization of the sublattices (in particular, $M_\text{s}=350$~kA/m for IrMn at room temperature). The dynamics of the N\'eel vector ${\bm l}(t)$ is governed by the well-known equation of the so-called  "sigma-model" ~\cite{khymyn2017antiferromagnetic1, satoh2010spin, khymyn2016transformation, SnchezTejerina2020}
\begin{equation}
{\bm l}\times \left[\frac{1}{\omega_\text{ex}}\frac{d^2\bm{l}}{dt^2} + \alpha_{\text{eff}} \frac{d\bm{l}}{dt} + \hat{\Omega}\cdot{\bm l} + [{\bm \tau}\times {\bm l}]  \right] = 0.
\end{equation}

Here  $\alpha_\text{eff}$ is the effective Gilbert damping constant ,${\bm e}_\text{e}~=~\bm{e}_\text{x}$ is the easy \textcolor{red}{a}xis of the AFM anisotropy, $\hat{\Omega}=~-~\omega_\text{e}~{\bm e}_\text{e}~\otimes~{\bm e}_\text{e}$,  $\tau~=~(\omega_\text{DC} \bm{n}_\text{DC}~+~\omega_\text{AC}\bm{n}_\text{AC}e^{i\omega t}~+~\text{c.c.})$ is the DC and AC spin-transfer torque intensity. Characteristic frequencies are defined as follows:
\begin{equation}
\left(\begin{array}{ccc} \omega_\text{DC} \\ \omega_\text{AC} \end{array}\right)=\sigma\cdot \left(\begin{array}{ccc} j_\text{DC} \\ j_\text{AC} \end{array}\right), \left(\begin{array}{ccc} \omega_\text{ex} \\ \omega_\text{e} \end{array}\right)=\gamma\cdot \left(\begin{array}{ccc} H_\text{ex} \\ H_\text{e} \end{array}\right),
\end{equation}
 where $j_\text{DC}$ and $j_\text{AC}$ are the densities of the input DC electric current and the AC electric current having frequency $\omega$, respectively, $\gamma$ is the modulus of the gyromagnetic ratio, $H_\text{ex}$ is the AFM internal exchange magnetic field and $H_\text{e}$ is the AFM anisotropy field. The torque-current proportionally coefficient $\sigma$ is determined by the following expression ~\cite{khymyn2017antiferromagnetic1}:
 \begin{equation}
 \sigma~=~\frac{e \gamma \theta_\text{SH} g_{r} \rho \lambda_\text{Pt}}{2\pi M_s d_\text{AFM}} \tanh\left( \frac{d_\text{Pt}}{2\lambda_\text{Pt}} \right),
 \end{equation}
 where  $g_{r}~=~6.9~\cdot~10^{18}$ m$^{-2}$ is the spin-mixing conductance at the AFM-Pt interface,  $d_\text{AFM}$ is the thicknesses of the AFM layer, $\theta_\text{SH}=0.1$ is the spin-Hall angle in Pt, $\rho~=~4.8~\cdot~10^{-7}\Omega~\cdot~$m is the electrical resistivity of the Pt layer, $\lambda_\text{Pt}=7.3$nm is the spin-diffusion length in Pt, $d_\text{Pt}=20$nm is the Pt thicknesses. In our numerical simulations, we used a typical value of the effective Gilbert damping constant for IrMn ~\cite{Gomonay2015, Zhang2019} $\alpha_{\text{eff}}~=~0.005$, which gives the quality factor $Q$ of the AFMR resonance approximately equal to \textcolor{red}{7}.
 
 The resonance frequency of the detector in a simplest case of a uniaxial easy-axis AFM can be calculated using the expression ~\cite{khymyn2017antiferromagnetic}:
\begin{equation}
\omega_\text{AFMR}=\sqrt{\omega_\text{ex}\omega_\text{e}},
\end{equation}
 where $\omega_\text{ex}$ is the exchange frequency and $\omega_\text{e}$ is the anisotropy frequency. For the uniaxial IrMn, with $\omega_\text{ex}/2\pi=12.9$~THz and $\omega_\text{e}/2\pi=16$~GHz, the AFMR frequency $\omega_\text{AFMR}/2\pi=454$~GHz \textcolor{red}{~\cite{Gomonay2015}}.

   \begin{figure}
	\includegraphics[width=0.45\textwidth]{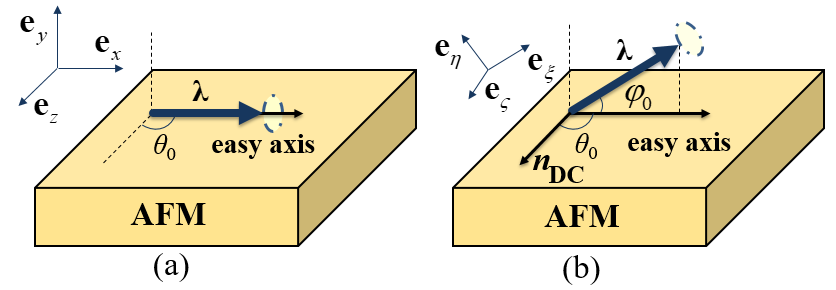}%
	\caption{\label{Fig_2} Schematic representation of a N\'eel's ground state vector $\bm{\lambda}$ orientation in an AFM for $j_\text{DC}=0$ (a) and  for $j_\text{DC}\ne 0$ (b).}%
\end{figure}

We can describe the small-amplitude dynamics of the N\'eel vector as ${\bm l}~=~\bm{\lambda}+\bm{s}e^{i\omega t} +\text{c.c.}$, where $\bm{\lambda}$ is the ground state of the N\'eel vector, while vector $\bm{s}$ describes the excitation created by the external AC spin current. These vectors satisfy the  orthogonality condition $(\bm{\lambda}\cdot\bm{s})=0$. 

Let us, first consider the situation when the bias DC current is absent. The ground state orientation of the N\'eel vector for the zero input DC current density is ${\bm \lambda}=(1,0,0)$ (see Fig.2a), so vector $~\bm{\lambda}$ is oriented along the easy axis. The oscillations of the dynamic vector $\bm{s}$ occur in the $(\bm{e}_\text{z,y})$ plane, and the projection $s_x=0$. In this case after simplifications we can find an expressions for the vector $\bm{s}$ in the following form:
\begin{equation}
\bm{s}= \frac{\omega_\text{AC} \omega_\text{ex}}{\omega^2_{AFMR}-\omega^2+i\gamma_0\omega} \cdot \bm{n}_\text{AC},
\end{equation}
where $\gamma_0=\alpha_{\text{eff}}\omega_\text{ex}$ is the spectral linewidth of the AFMR for the zero input DC current.

The rectified output DC spin current in the top Pt layer (see Fig.1) is proportional to  $\bm{j}_\text{out}\sim  [\bm{l}\times\dot{\bm{l}}]$ ~\cite{khymyn2017antiferromagnetic1, khymyn2017antiferromagnetic}. From Eq.(5), after simplifications, one can find the output DC electric voltage between output contacts in the following form~\cite{khymyn2017antiferromagnetic}:
\begin{equation}
V_\text{out}(\omega) = \frac{V^0_\text{max}(\gamma_0)^2\omega_{AFMR}\omega}{(\omega_{AFMR}^2-\omega^2)^2+(\gamma_0\omega)^2}.
\end{equation}
Here $V^0_\text{max}=\left( \frac{\omega_\text{AC}}{\gamma_0}\right)^2 V_0$ is the maximum output DC voltage in the resonance case $\omega=\omega_{AFMR}$, and the normalized voltage $V_0$ is defined by the expression\cite{khymyn2017antiferromagnetic1}:
\begin{equation}
V_0=L~e~\omega_\text{ex}^2~\theta_\text{SH}~g_r~\rho~\lambda_\text{Pt}~\frac{\tanh(d_\text{Pt}/2\lambda_\text{Pt})}{\pi d_\text{Pt} \omega_\text{AFMR}}.
\end{equation}

In our calculations we assume that the AFM layer has a square cross section with the in-plane dimensions $S=100\times 100$ nm$^2$ and the thickness $d_\text{AFM}$ is 5 nm. For the $j_\text{AC}~=~10^7$A/cm$^2$ the output electric voltage to be $V_\text{out}~\approxeq~100 \mu$V at the zero DC input current density.

Fig.3 shows the standard resonance-type dependence of the output voltage $V_\text{out}$ on the frequency $\omega$ of the input AC signal. As it can be seen from Fig.3 the spectral linewidth of the output voltage is equal to the linewidth of the AFM resonance $\gamma_0/2\pi=64.8$ GHz.

   \begin{figure}
	\includegraphics[width=0.45\textwidth]{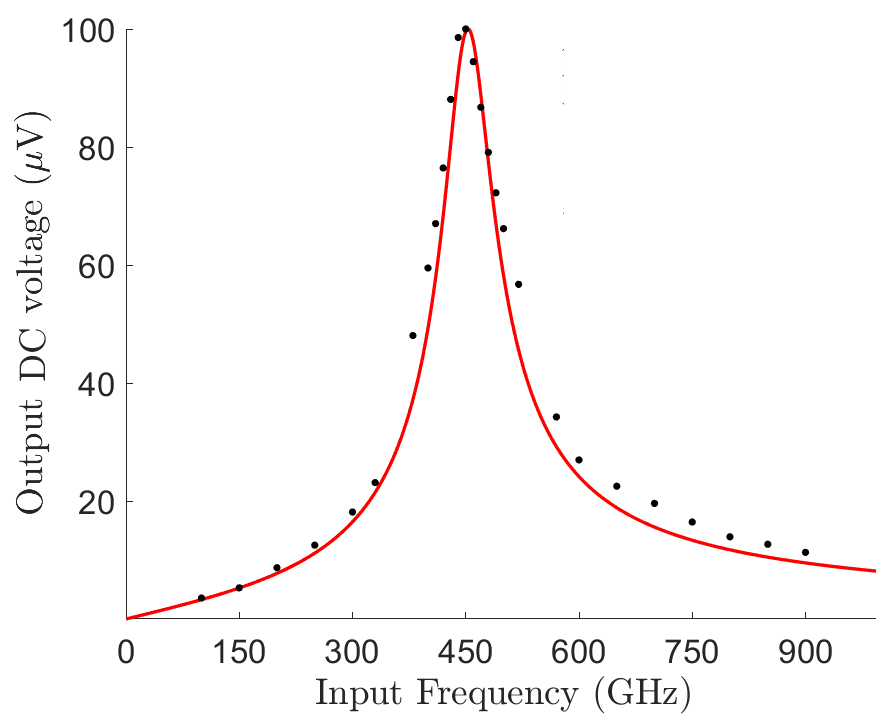}%
	\caption{\label{Fig_3} Resonance curve of the output rectified DC voltage $V_{out}$ of an AFM-based detector calculated using Eq.(6)(solid line). Dots show the results of micromagnetic simulations.}
\end{figure}

Now, let us consider the case when a non-zero bias input DC-current is applied to the detector Fig.1. In this case (see Fig.2b) the static equation defining the ground state Neel vector $\bm{\lambda}$ can be found from (1), and  has the following form:
\begin{equation}
\bm{\lambda}\times (\hat{\Omega}\cdot \bm{\lambda}) + \omega_\text{DC}\bm{\lambda}\times \bm{n}_\text{DC}\times\bm{\lambda}=0.
\end{equation}

Using a spherical coordinate system, one can express the ground state of the N\'eel vector as: $\bm{\lambda}~=~(\cos(\varphi_0)\sin(\theta_0),\sin(\varphi_0)\sin(\theta_0),\cos(\theta_0))$. It also follows from (1) that for $\bm{n}_\text{DC}={\bm{e}_\text{z}}$ and  $\bm{n}_\text{e}=\bm{e}_\text{x}$ the azimuthal angle $\theta_0\approxeq \pi/2$, and the static polar angle of the Neel vector $\bm{\lambda}$ is:
\begin{equation}
\varphi_0=\frac{1}{2}\arcsin\left(\frac{2\omega_\text{DC}}{\omega_\text{e}}\right).
\end{equation}

The increase of the polar angle $\varphi_0$ means that the static part of the N\'eel vector is deflected from the plane of the AFM interface, which results from the spin-transfer-torque induced by the spins injected from the bottom Pt layer traversed by the external bias DC current.

The "dynamic" equations defining the "excitation" vector $\bm{s}$, after some simplifications, can be written in the following form:
\begin{gather}
\left(-\frac{\omega^2}{\omega_\text{ex}} +i\omega\alpha_{\text{eff}}\right)\bm{s} + \left( \hat{\Omega} -\left(\bm{\lambda} \cdot (\hat{\Omega}\cdot \bm{\lambda}) \right) \hat{I}\right)\cdot\bm{s} - \nonumber \\
-\left( \bm{\lambda} \cdot \left( \hat{\Omega}\cdot \bm{s}\right)\right)\bm{\lambda} + \omega_\text{DC}\left(\bm{\lambda}\cdot \bm{n}_\text{DC} \right)\bm{\lambda} \times \bm{s}=
\omega_\text{AC} \bm{\lambda} \times \bm{n}_\text{AC}.
\end{gather}

As it was mentioned earlier, for the zero input bias DC current the oscillations of the  vector $\bm{s}$ take place in the $(\bm{z,y})$ plane and $s_x=0$, whereas in the presence of the DC bias current, it is necessary to introduce a new coordinate system $(\bm{e}_{\xi},\bm{e}_{\eta},\bm{e}_{\zeta})$ (see Fig.2b), where the component $s_{\xi}$ perpendicular to the plane of oscillation ($\bm{e}_{\eta},\bm{e}_{\zeta}$) is equal to zero.  The expressions for the components $s_{\eta},s_{\zeta}$ are equivalent to those for the components $s_{y},s_{z}$ (5) if we replace $\bm{e}_{y}$ by $\bm{e}_{\eta}$, $\bm{e}_{z}$ by $\bm{e}_{\zeta}$, and  $\omega_{AFMR}$ by $\omega_{0}~=~\omega_\text{AFMR}~\sqrt{\cos(2\varphi_0)}$, respectively.

Recently~\cite{Popov2019} for the case of a biaxial AFM (e.g. NiO) it has been shown that in NiO there are two resonance frequencies $\omega_{1,2}$ , and they are substantially different due to the strong difference between the anisotropies corresponding to the "easy" and "hard" axes.
 In the case of a uniaxial AFM two AFMR eigenfrequencies are degenerate for the zero DC bias current. Although for a nonzero DC bias current the dependencies $\omega_{1,2}(j_\text{DC})$ differ, the difference is rather small, and in a uniaxial case we still can approximately assume, that $\omega_1\approxeq\omega_2=\omega_{0}$. The dependence of the resonance oscillation frequency on the DC bias current density in a uniaxial AFM  can be expressed as:
\begin{equation}
\omega_0 = \omega_\text{AFMR} \sqrt[4]{1-\left( \frac{2\omega_\text{DC}}{\omega_\text{e}} \right)^2}.
\end{equation}
The oscillation frequency (11) is  proportional to the AFMR frequency $\omega_\text{AFMR}$, and depends on the density $j_\text{DC}$ of the DC bias current.  Thus, the resonance frequency of magnetization oscillations in a uniaxial AFM can be tuned (reduced) by the variation of the bias DC current density $j_\text{DC}$ in the  bottom Pt layer of the detector Fig.1.

   \begin{figure}
	\includegraphics[width=0.45\textwidth]{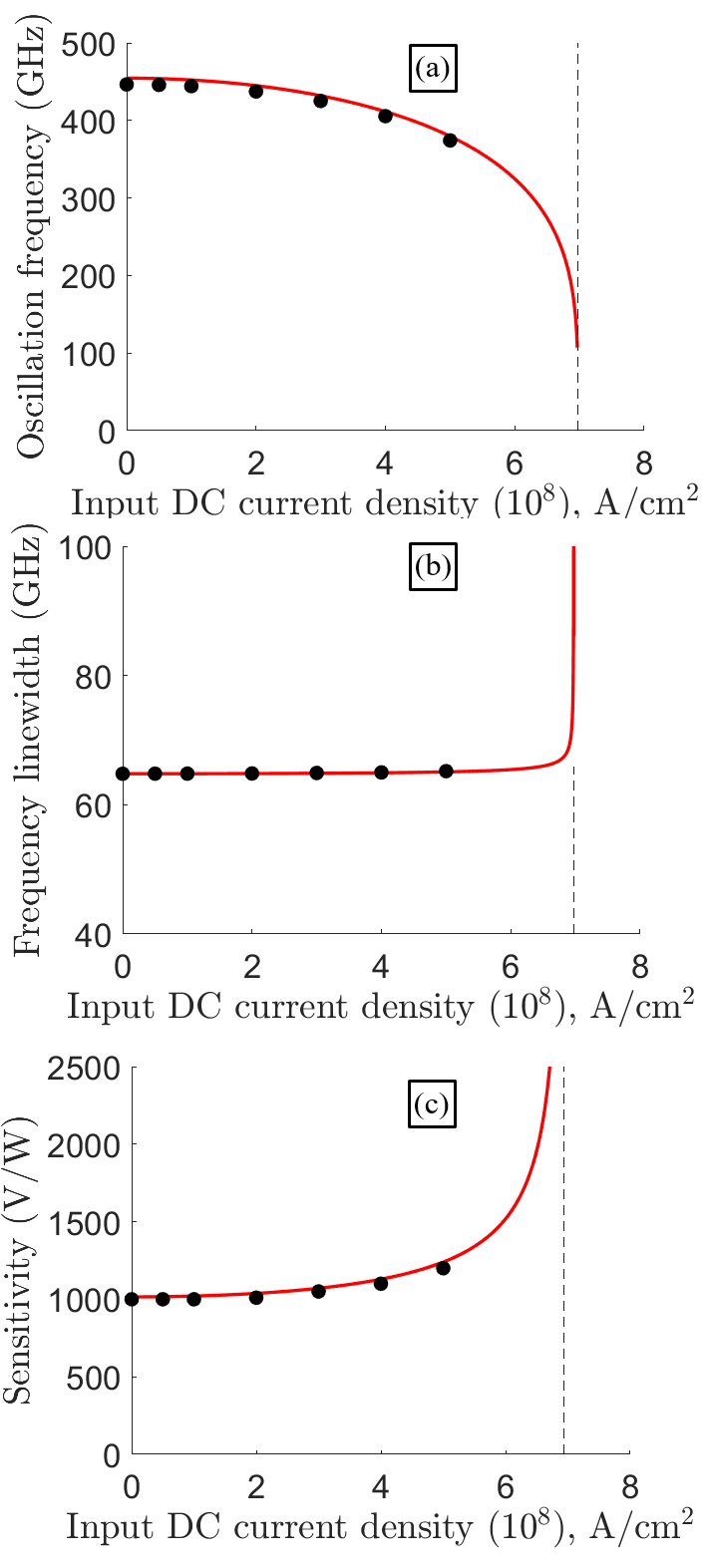}%
	\caption{\label{Fig_4} Comparison between the results of analytical calculations (solid line) and micromagnetic simulations (dots) showing the dependence of the AFMR oscillation frequency  (11) (a), frequency bandwidth (12) (b) and detector sensitivity (13) (c) on the input bias DC current density flowing in the bottom Pt layer of the detector Fig.1. Dashed lines show the dependences of the oscillation frequency, spectral linewidth and sensitivity in the strongly nonlinear regime near the threshold of the auto-oscillation (super-critical) regime. Dots show the results of micromagnetic simulations.}%
\end{figure}

Fig.4a shows the dependence of the oscillation frequency $\omega_0$ Eq.(11) as a function of the DC current density in the sub-critical (passive) regime. It is clear, that the frequency $\omega_0$ can be continuously reduced from $\omega_\text{AFMR}$ by at least 10 percent through the  increase of the bias current density to $ 5\cdot 10^8$ A/cm$^2$.

The threshold current $j_\text{th}$ at which the detector Fig.1 enters the auto-oscillation (super-critical or active) regime can be easily found from the stability analysis of the damped oscillation mode in Eq (1) in the form $j_\text{th} = \omega_\text{e}/(2\sigma)\approxeq 7\cdot 10^8$ A/cm$^2$ ~\cite{khymyn2017antiferromagnetic}. This value of the current density is rather high. Note, however, that both the AFMR frequency and the threshold DC current density corresponding to the transfer to the auto-oscillation regime can be substantially reduced in a PZ/AFM/Pt hetero-structures \cite{Popov2019} based on a thin dielectric AFM layer using the magneto-elastic interaction  for the voltage control of the AFM anisotropy.

 It can be shown from (5) that the dependence $\Delta \omega(j_\text{DC})$ for sufficiently low DC current densities can be found as follows:

\begin{equation}
\Delta\omega = \gamma_0 \left( 1+\frac{\alpha_{\text{eff}}^2 \omega_\text{ex}}{2\omega_\text{e}}\cdot \left(\frac{\omega_\text{DC}}{\omega_\text{e}}  \right)^2 \right).
\end{equation}
Fig.4b shows the dependence of the frequency linewidth $\Delta \omega$  on the input bias DC current density. As it can be seen from Fig.4b, the linewidth  in the sub-critical (passive) regime is practically independent of the DC bias current. Of course, for the larger DC bias current densities, close to the auto-oscillation threshold  the formula (12) is incorrect , and a more accurate theoretical analysis is necessary.  The linewidth 64.8 GHz achieved at the zero DC current density corresponds to the quality factor $Q=\omega_{0}/\Delta\omega\approxeq 7$.  Although this Q-factor is relatively small, it still is sufficient for the resonance reception of THz-frequency AC signals.

Finally, we calculate the detector sensitivity $\mathcal{R} = V_\text{max}/P_\text{AC}$, where $V_\text{max}$ is the maximum rectified DC voltage $V_\text{max}=V^0_\text{max}\cdot\frac{\cos(\varphi_0)}{\sqrt{\cos(2\varphi_0)}}$ for the non-zero input bias DC current, $P_\text{AC} = \rho j_\text{AC}^2 S d_\text{AFM}$ is the input AC electric power. For the $j_\text{AC}~=~10^7$A/cm$^2$ the input AC power is $P_\text{AC}~\approxeq~100$ nW. The output signal (see Fig.1) from the Pt layer can be detected via the ISHE in the symmetric Pt/AFM/Pt structure (see for more details e.g.\cite{Yang2016, Wu2016}).

At the signal frequencies close to the AFMR the detector sensitivity $\mathcal{R}_0$ exceeds 1000 V/W.  Fig.4c shows the dependence of the resonance sensitivity $\mathcal{R}$ on the input DC bias  current density, which is calculated using Eq. (6) for $\omega=\omega_{0}$ and relatively small DC bias current  densities in the following form:

\begin{equation}
\mathcal{R} = \mathcal{R}_0 \cdot \left( 1+\frac{1}{2}\cdot \left(\frac{\omega_\text{DC}}{\omega_\text{e}}\right)^4\right).
\end{equation}
Here $\mathcal{R}_0=\left( \frac{\sigma}{\gamma_0} \right)^2 \frac{V_0}{\rho S d_\text{AFM}}$ is the sensitivity for the zero DC bias current density. As it can be seen from (13), the sensitivity increases slightly with the increase of the input DC bias current density. \textcolor{red}{Note that the more detailed calculation of the sensitivity should be made with taking into account thermal fluctuations in the AFM~\cite{Semenov2019}, which is the subject of the separate work.}

 In conclusion, we have demonstrated theoretically that an AFM having uniaxial anisotropy can be used as a sensitive element for the resonance detection of THz-frequency spin currents. We have shown that an additional bias DC current in the HM layer can be used to reduce  the effective anisotropy of the AFM layer, and, therefore, to continuously tune the  AFM resonance frequency. Analogous calculations can be made for a biaxial AFM (NiO, hematite $\alpha$-Fe$_2$O$_3$, etc.) as well. The proposed AFM/HM hetero-structure works as a resonance-type quadratic detector which can be tuned by a bias DC current in the range of at least 10 percent of the AFMR frequency. We have shown that for the zero input DC current, a circularly polarized external AC current excites the rotation of the Neel vector in a plane perpendicular to the interface. Practical realization of circularly polarization of the AC current is the subject of a separate study and it will be considered in our future works. Our estimations also show that the sensitivity of the proposed AFM detector of THz-frequency signals could be comparable to that of modern detectors based on the Schottky, Gunn or graphene-based diodes ~\cite{lewis2014review} (with maximum sensitivity of the order of $10^5$ V/W). We anticipate that this described detector effect can be  observed experimentally using the electric injection and detection of spin currents via the spin Hall effects in HM/AFM bilayers, like in the pioneering experiments shown in~\cite{lebrun2018tunable, Tarequzzaman2018}. In our opinion, the proposed AFM-based resonance detectors of THz-frequency signals can be used as general-purpose receivers of THz-frequency radiation, as sensors for the detection of THz-frequency spin currents generated by ultra-fast AFM artificial neurons in neuromorphic computational architectures ~\cite{khymyn2018ultra, sulymenko2018ultra}, and as sensitive elements in THz-frequency spectrum analyzers~\cite{Artemchuk2020}, and CMOS-compatible AFM-based memory cells~\cite{Shi2020}.

This work was supported in part by the U.S. National Science Foundation (Grants No. EFMA-1641989), by the U.S. Air Force Office of Scientific Research under the MURI grant No. FA9550-19-1-0307, and by the Oakland University Foundation.

The work of V.P., M.C. and G.F. was supported under the Grant 2019-1-U.0. ("Diodi spintronici rad-hard ad elevata sensitività - DIOSPIN") funded by the Italian Space Agency (ASI) within the call “Nuove idee per la componentistica spaziale del futuro”.

We would like to acknowledge networking support by the COST Action No. CA17123 “Ultrafast opto magneto electronics for non-dissipative information technology”.

Studies of the detection were carried out with the support of Russian Science Foundation (Grant No. 19-19-00607) and the grant from the Government of the Russian Federation for state support of scientific research conducted under the guidance of leading scientists in Russian higher education institutions, research institutions and state research centers of the Russian Federation (Project No.075-15-2019-1874). Studies of the magnetic dynamics in antiferromagnetic were supported by the RFBR under the grants Nos. 18-37-20048, 18-29-27018, 18-57-76001, 18-07-00509 A, 18-29-27020, and by the grants of the President of the Russian Federation (No. MK-283.2019.8, No. MK-3607.2019.9). S.A. acknowledges support from the Government of the Russian Federation (grants No. 074-02-2018-286 for the ”Terahertz Spintronics” laboratory of the MIPT). A.R. acknowledges the support from the RFBR (grant No. 19-29-03015). P.A. achnowledges the support from the RFBR (grant No. 19-32-90242).

\subsection*{DATA AVAILABILITY}
The data that support the findings of this study are available from the corresponding author upon reasonable request.

\subsection*{REFERENCES} 

\bibliography{aipsamp}

\end{document}